# Dark Rate of a Photomultiplier at Cryogenic Temperatures

H.O. Meyer

*Physics Dept., Indiana University, Bloomington, IN 47405*
*(meyer1@indiana.edu)*

**Abstract.** When cooled below room temperature, the pulse response of a photomultiplier in the absence of light (dark rate) initially decreases, but then turns around near 250 K and continues to rise all the way down to 4 K. When the photomultiplier is cold, its dark response is burst-like. We have measured the characteristics of the dark response of a photomultiplier.

## INTRODUCTION

Dark counts are the response of photonic detectors in the absence of light. In a photomultiplier (PM) at room temperature, thermionic emission of electrons is the major source of dark counts. When the PM is cooled, the thermionic dark rate decreases with temperature according to Richardson's law.

In 1963, while cooling a PM with dry ice, Rodman and Smith [1] noticed that below about 250 K, the observed dark rate was larger than expected from thermionic emission. They coined the term 'non-thermal' dark rate for this effect. Additional sources for dark counts include cosmic-ray Čerenkov radiation in glass envelope, radioactivity in the glass or the environment, and field emission of electrons. However, as will become clear, none of these effects explains the dark counts in a cold PM and thus a new mechanism seems to be at work.

Recently, interest has been growing in the use of photomultipliers for observing scintillation in noble liquids in experiments concerned with neutrino interactions, dark matter searches, the electric dipole moment of the neutron etc. One of the obstacles that had to be overcome was the loss of conductivity of the photocathode material when cooled. This problem, which prevents the operation of the PM below about 150 K, was cured by depositing the cathode onto a thin platinum layer. Consequently, a number of such PM tubes have been tested for cryogenic application. During these tests it became apparent that the non-thermal dark rate, the onset of which has been seen earlier, is in fact *increasing* with falling temperature [2,3].

In this paper we present a dedicated study of the dark response of a cold photomultiplier.

## EXPERIMENT

The photomultipliers used in this experiment are model R7725 tubes, manufactured by Hamatsu. These tubes are 5cm in diameter with a K-Cs-Sb (bialkali) photocathode (area $A_c$ = 17 cm$^2$) with a Pt underlayer, and 12 dynode stages. At an operating voltage of 1800 V the gain is typically $10^8$.

The PM under test is housed in a sealed, cylindrical stainless steel container, which is hanging by three G10 rods from a support flange at the top of a bucket Dewar. A corrugated thin-walled stainless steel pipe connects the room-temperature top to the test container at the bottom of the Dewar. The pipe serves as a conduit for the high-voltage supply wires, the signal cable, and thermometer connections. It is also used to evacuate the container. When the Dewar is filled with liquid nitrogen or helium, the liquid comes into contact with the outer container wall. Cooling of the PM takes place by radiation between it and the container wall. The inside wall of the container is painted black and the PM is wrapped in black tape to increase emissivity and thus the cooling rate. The temperature $T$ is measured on the cylindrical glass surface of the PM, next to the photocathode. The fastest cooling rate observed is about 30 K/hour. All vacuum feedthroughs are at the top flange, at room temperature. The voltage divider, located outside the vacuum, is connected to individual dynodes by a custom-made, low-conductivity ribbon cable.

The anode signal is amplified and discriminated at about ⅓ of the amplitude of a single-electron event. The ‘dark events’ (discriminator pulses) are counted by a programmable scaler. In addition, the time intervals $\Delta t_n$ between subsequent pulses are measured to within 1 μs with a simple CAMAC setup involving a 1 MHz clock pulse generator. The readout of the pulse time takes about 4 μs; events that occur within this dead time are also counted.

## RESULTS

### Average Dark Rate

After applying voltage, the dark rate changes as the tube is conditioning. Typically, the dark rate decreases by about a factor of five over a period of a day or two. The room-temperature dark rate of a conditioned PM varies greatly from tube to tube. For instance, for five tubes of the same model (operated at the same gains) the dark rates vary by about a factor of 40. At room temperature, the dark rate also depends on the operating voltage, as will be discussed later.

When the PM is cooled, the dark rate *r(T)* initially decreases in accordance with Richardson's law for thermionic emission

$$r_R(T) \propto T^2 \exp\left(-\frac{W}{kT}\right). \tag{1}$$

This behavior is shown in fig.1, by the dashed lines that are calculated from eq.1 with $W$ = 0.5 eV, and arbitrarily normalized to the data.

Fig.1 shows a measurement of the dark rate *r(T)* for two tubes of the same model (R7725). Both tubes are modified with a platinum underlayer. The first tube (triangles) is studied all the way down to liquid-helium temperature. The second tube is cooled with liquid nitrogen (squares), and then warmed up again (diamonds). The hysteresis effect between 100 K and 200 K is an artifact due to the slight delay between the measured and the actual cathode temperature.

The dark rates *r(T)* for the two tubes differ at room temperature by an order of magnitude, however, below 200 K both tubes follow a common exponential dependence. Our data may be compared with a similar measurement for a Hamamatsu R5912 tube, carried out by a group at Yale University [2]. They find a temperature dependence of the dark rate that is very similar to the one shown here, but at an overall rate that is higher by a factor of 20. It is interesting to note that the cathode area of the R5912 ($A_c$ = 335 cm$^2$) is also 20 times larger than that of our R7725 ($A_c$ = 17 cm$^2$). This suggests that the dark rate scales with the cathode surface area, and that the following 'universal' expression for the non-thermal dark rate is valid for $T$ < 200 K

$$r_{nt}(T) = G\,A_c \exp\left(-\frac{T}{T_r}\right) \ , \tag{2}$$

where $A_c$ is the cathode area, $G$ = (5±1) Hz/cm$^2$ is an empirical constant, and $T_r$ = 100 K has been adjusted to the slope of the data in fig.1 below 200 K. The solid line in fig.1 is calculated with eq.2. A further test of eq.2 is provided by the reported [3] dark rate of (1142±235) Hz, measured at 77 K for an ETL 9357FLA tube ($A_c$ = 430 cm$^2$). This datum is also in agreement with eq.2. We need to emphasize that all the data that are used to support eq.2 have been acquired with tubes with bialkali cathodes on a platinum backing.

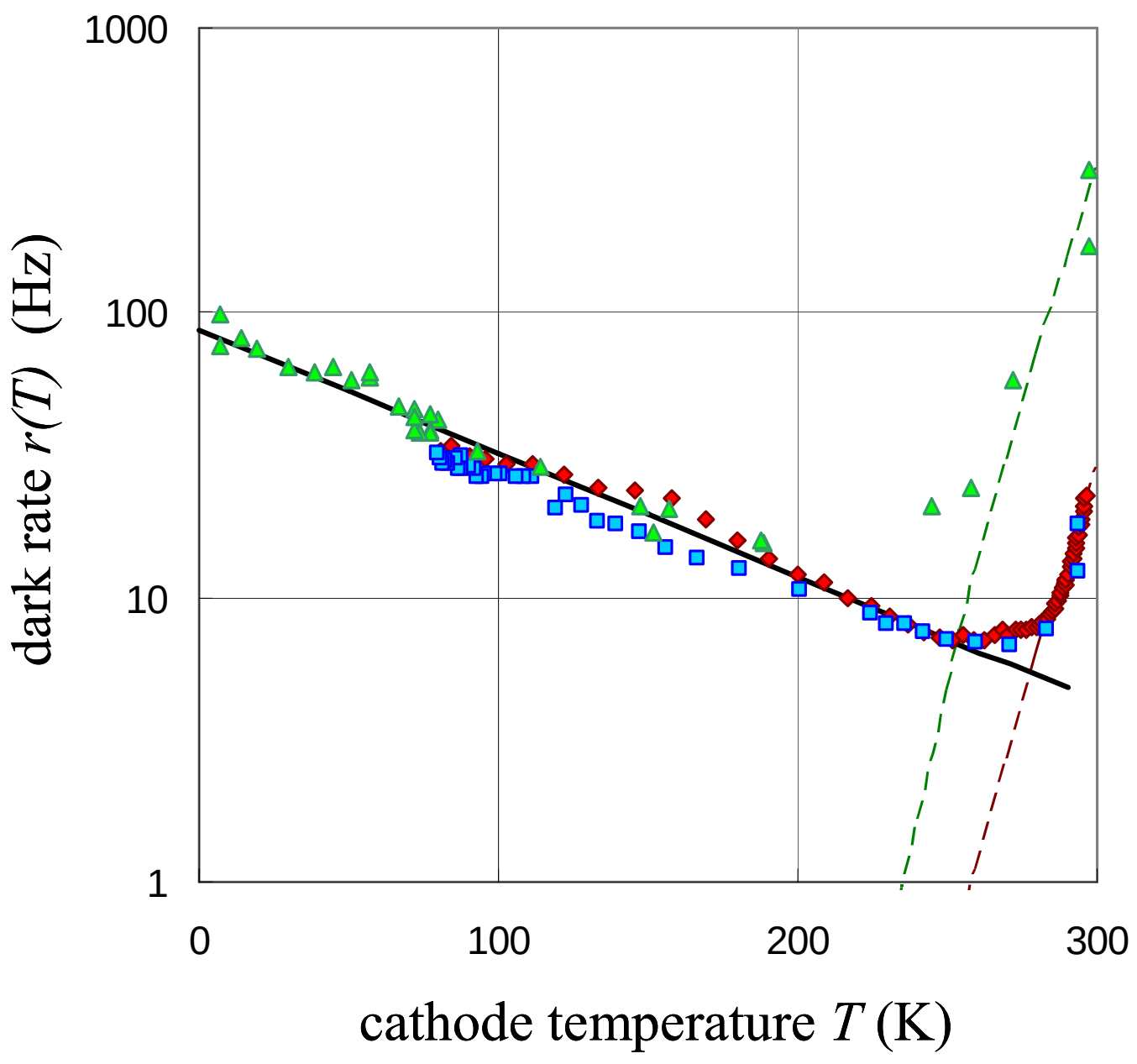

**FIGURE 1.** Dark rate for two tubes of the same model (R7725). Tube 1 was cooled to 4 K (triangles). Tube 2 was cooled to 80 K (squares) and then warmed back up to room temperature (diamonds). The solid line corresponds to eq.2. The dashed curves indicate the temperature dependence expected for thermionic emission.

The non-thermal dark effect was discovered [1,4] as the departure of the data from Richardson's law. The cathode materials in these cases were Sb-Cs and Sb-Na-K-Cs, respectively. In addition, we have also studied a 'normal' R7725 tube, with a K-Cs-Sb cathode but *without* platinum backing. These tubes do not allow a measurement below about 150 K, but they all clearly show the onset of the non-thermal dark effect. One thus can state that the non-thermal dark effect also occurs in tubes without platinum underlayer, and with (at least three) different cathode materials.

The amplitudes of non-thermal dark events have been measured and found to be mostly consistent with that of single-electron events. These electrons originate at the photocathode (and not the dynodes) since it has been shown that the dark rate goes to zero when the cathode potential is set to that of the first dynode [1].

Fig.2 shows the dependence of the dark rate on the operating voltage $U_{PM}$ of the PM. At room temperature (solid symbols), the dark rate depends on $U_{PM}$, as one would expect since the electrical field at the cathode surface is changing, affecting thermionic emission (Schottky effect). However, the non-thermal dark rate (open symbols) shows no such dependence. Thus, non-thermal electron emission is not significantly affected by the electrical field at the cathode surface.

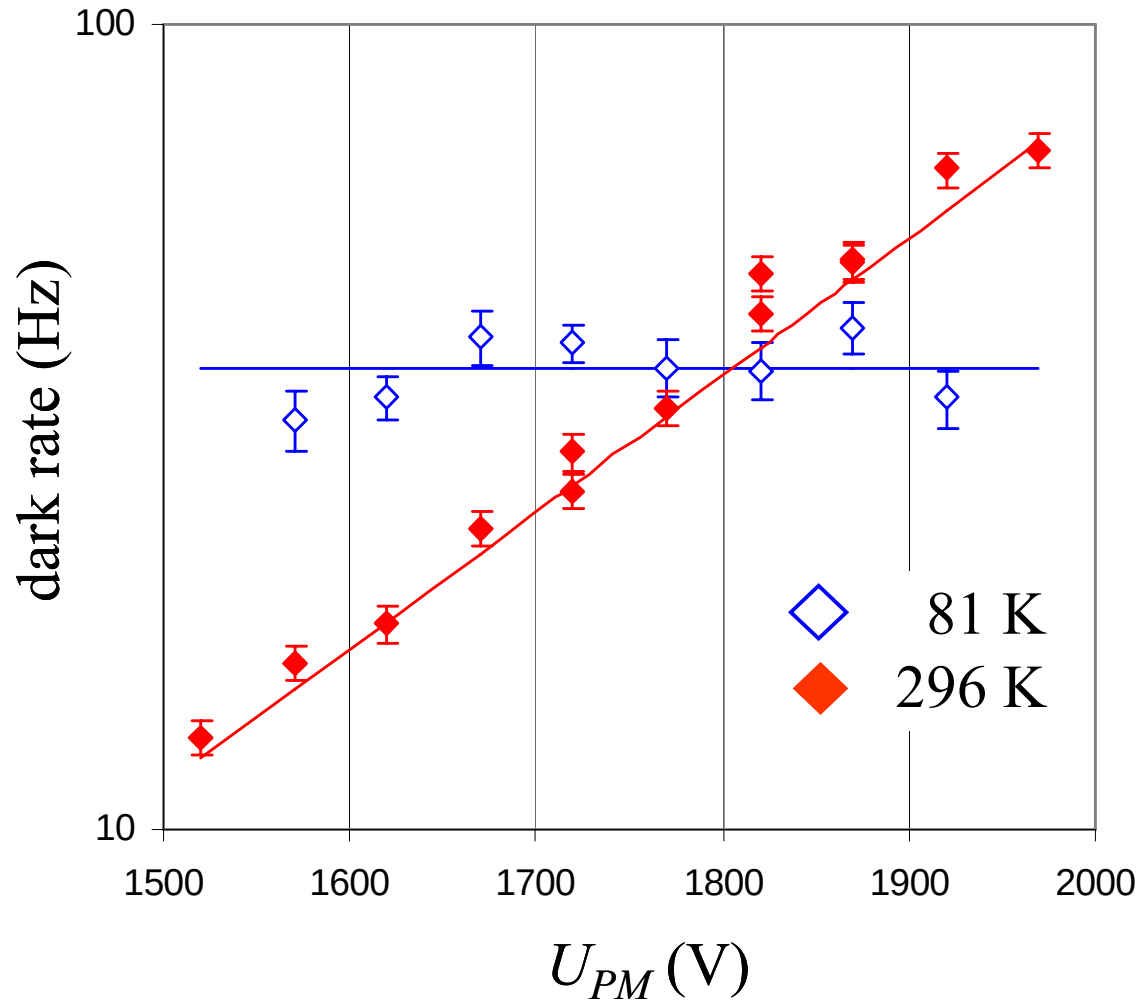

**FIGURE 2.** Dark rate as a function of the operating voltage of the photomultiplier.

## Time Distribution of Dark Events

### *Distribution of Spacings between Events*

Already Rodman and Smith [1] have suspected that the dark rate is 'nonstatistical in character'. In order to investigate this, we study the time distribution of dark pulses at liquid-nitrogen temperature. To this effect we measure, for a run of $N$ pulses, the time $\Delta t_n$ elapsed since the previous pulse ($n$ = 1…$N$). The sum of all $\Delta t_n$ is then the duration $t_N$ of the run. A typical run lasts about 15 minutes, and consists of 2 to 3 $10^4$ such time intervals.

The frequency distribution for a typical measurement at 81 K is shown in fig.3a. For random events (occurring independently of each other at an average rate $\rho$) the distribution of intervals $\Delta t$ would be proportional to $\exp(-\rho\Delta t)$. The data in fig.3a indeed exhibit such an exponential dependence (solid line), except for a spike in the first few bins, which indicates an excess of short intervals above the random expectation. This spike, in fact, contains the information that will be discussed in the remainder of this paper.

A closer look at the spike region is shown in fig.3b. The excess of intervals above the random distribution (solid line), comprise about 70% of all collected events ($N$). It may be worth noting that the distribution of these excess events, to a very good approximation, is proportional to $\Delta t^{-2/3}$ and thus follows a power law (dashed green line in fig.3b).

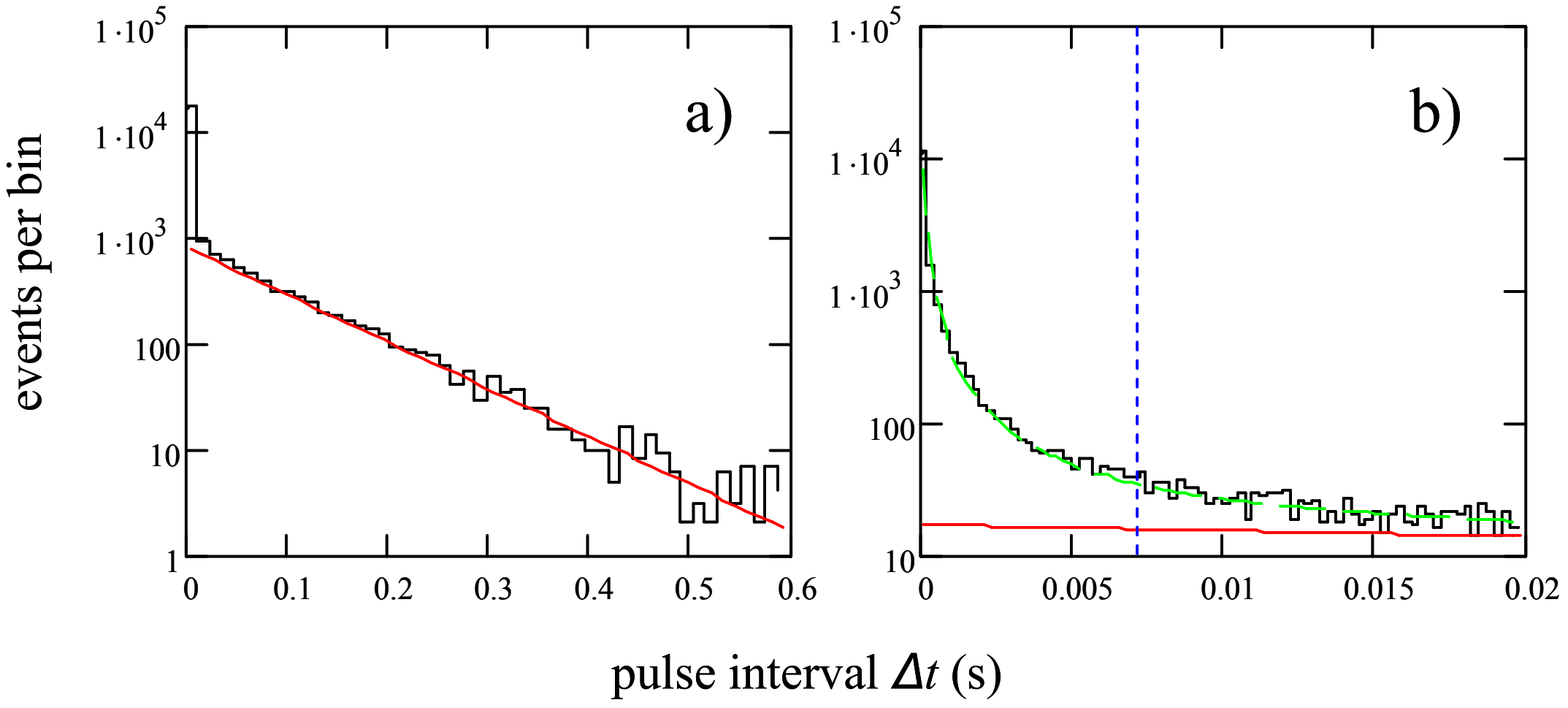

**FIGURE 3.** a) Distribution of $N = 30391$ intervals $\Delta t$ observed at $T = 81$ K. The time bins are 12 ms wide. The red solid line is an exponential fit, starting at 0.04 s.
.b) Part of fig. 1a on an expanded scale. The time bins are 250 µs wide. The dashed blue line divides the data into 'short' and 'long' intervals. There are $N_s = 21411$ short intervals ($\Delta t < 7$ms). The dashed green line is mentioned in the text.

In the following, we study how these excess events are grouped in time. We start by dividing the data into 'short' and 'long' intervals, by introducing a separator $\theta_{sep}$ (shown in fig. 3 as a vertical dashed line at $\Delta t = 7$ ms). Almost all (97%) of the 'short' events are excess events (above the solid line), and almost all excess events are to the left of the separator. This fact does not depend much on the choice of the separator nor do the conclusions that follow.

About 10 % of the dark pulses occur during the 4 µs dead time, while reading the clock scaler. These pulses are counted in a second scaler channel, and are inserted into the $\Delta t_n$ data sequence at the appropriate place with a nominal $\Delta t = 2$ µs. Thus, there are no events lost due to dead time.

## *Bursts and the Distribution of Burst Size*

We define a 'burst' as an uninterrupted sequence of 'short' intervals. In a data set of $N$ events, there are $N_s$ 'short' intervals and $M$ bursts. The 'size' $L$ of a burst is the number of intervals in the burst. The bursts are numbered with $m = 1 \ldots M$, and the size of the $m^{\text{th}}$ burst is $L_m$. The average burst size is given by $\langle L \rangle = N_s / M$.

The number of bursts of a given size $L$ shall be denoted by $F_L$. The number of intervals associated with all bursts of that size is $L \cdot F_L$. The latter quantity is plotted in fig.4 for a typical run at 81 K. Given the fact that each interval corresponds to an electron emitted from the cathode, the abscissa in fig.4 may be interpreted as the burst charge in units of the elementary charge $e$. The ordinate is then proportional to the cathode current that is associated with a given burst charge.

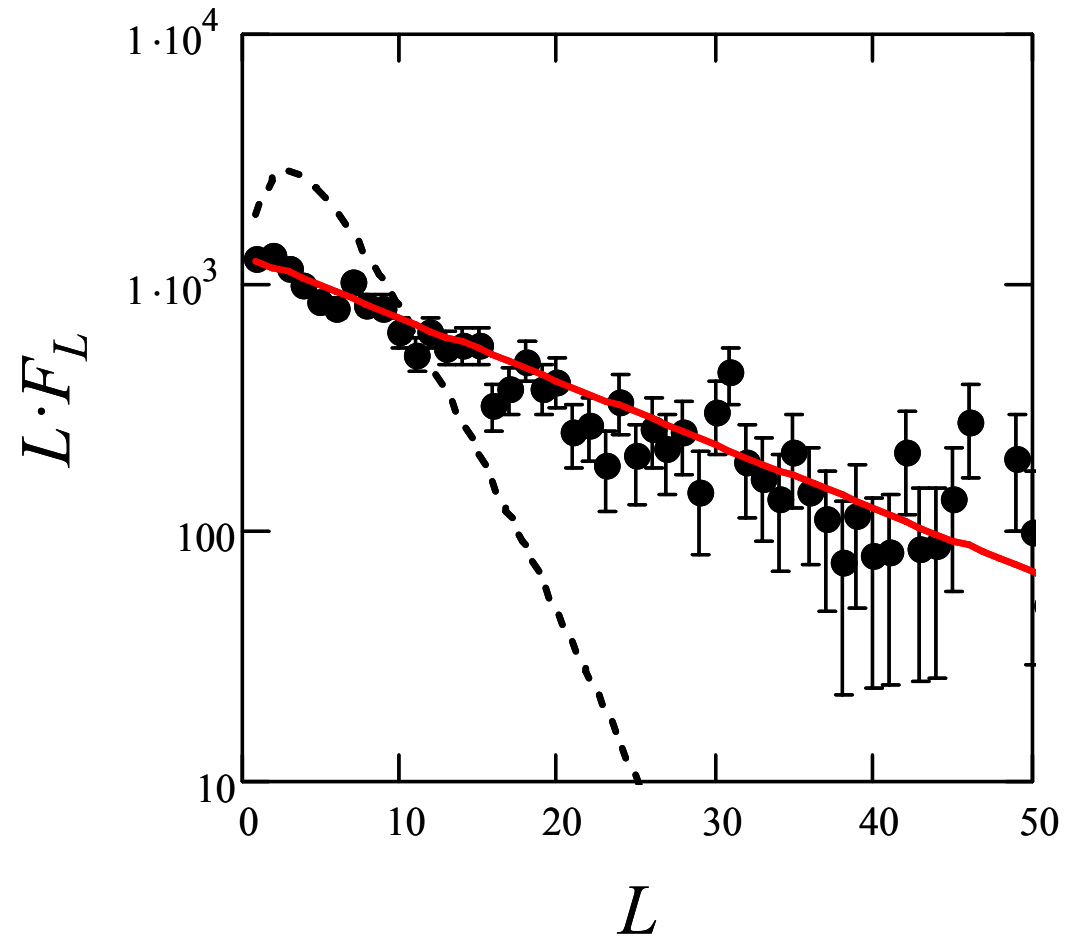

**FIGURE 4.** Number of intervals, $L \cdot F_L$, associated with bursts of size $L$, measured at 81 K. The total number of bursts is $M = 3711$, the average burst size is $\langle L \rangle = 5.77$. The solid line is calculated from eqs. 3 and 4, using these values. The dashed line shows the distribution that would result if the same number of short and long intervals were arranged at random. Using $p$=0.705 for the fraction of short intervals (see caption fig.3), one finds for the random distribution $M_{rand} = 6316$, $\langle L \rangle_{rand} = 3.38$.

The fact that the data in fig.4 follow an exponential distribution justifies the Ansatz

$$L \cdot F_L = N_s \, \frac{1-q}{q} \; q^L \quad , \tag{3}$$

where the constant $q$ is determined by the data, and $N_s$ is the number of short intervals. The normalization factor follows from the fact that the sum over all $L \cdot F_L$ equals $N_s$, where the sum of the geometric series is taken from $L = 1$ to infinity. Similarly, setting the sum over all $F_L$ to $M$, the total number of bursts, leads to the following relation between the parameter $q$ and the average burst size:

$$\langle L \rangle = -\frac{q}{(1-q)\,\ln(1-q)} \quad . \tag{4}$$

Therefore, the burst size distribution is completely determined by the average burst size $\langle L \rangle$ and the number of bursts $M = N_s / \langle L \rangle$. The solid line in fig.4 is obtained with eqs. 3 and 4, using the numbers given in the captions of figs. 3 and 4.

It is important to realize that bursts according to our definition occur even if the short and long intervals in the sequence are arranged at random. Using $p = N_s / N$ as the probability that a given interval is short, it is easy to calculate the burst size distribution that results from a random arrangement of the same fraction of short intervals as has been measured. The resulting distribution is shown as a dashed line in fig.4. For such a random sequence, the average burst size is given by $<L>_{rand} = (1-p)^{-1}$ and the number of bursts by $M_{rand} = N_s\,(1-p)$. Thus, for a fixed fraction $p$ of short intervals, any physical effect that causes an accumulation of short intervals has the effect of lowering the number of bursts and increasing their size with respect to their random values (numbers are given in the caption of fig.4).

The smallest burst ($L = 1$) is a pair of pulses separated by a short interval. Not included are single pulses, i.e., pulses that are preceded and followed by a long interval. The number of these 'bursts of size zero', is given by $F_0 = N - M(<L> +1)$, i.e., the number of events *not* associated with bursts. These single pulses are responsible for the exponential part in fig.3. The complete burst size distribution $F_L$ is shown in fig.5. Fig.5 is identical to fig.4, except for a factor $1/L$ on the ordinate and the inclusion of $F_0$ (open symbol). The latter is well explained by eq.3 (solid line). Thus, single events naturally fit into the burst size distribution $F_L$ ($L = 0, 1,\ldots$).

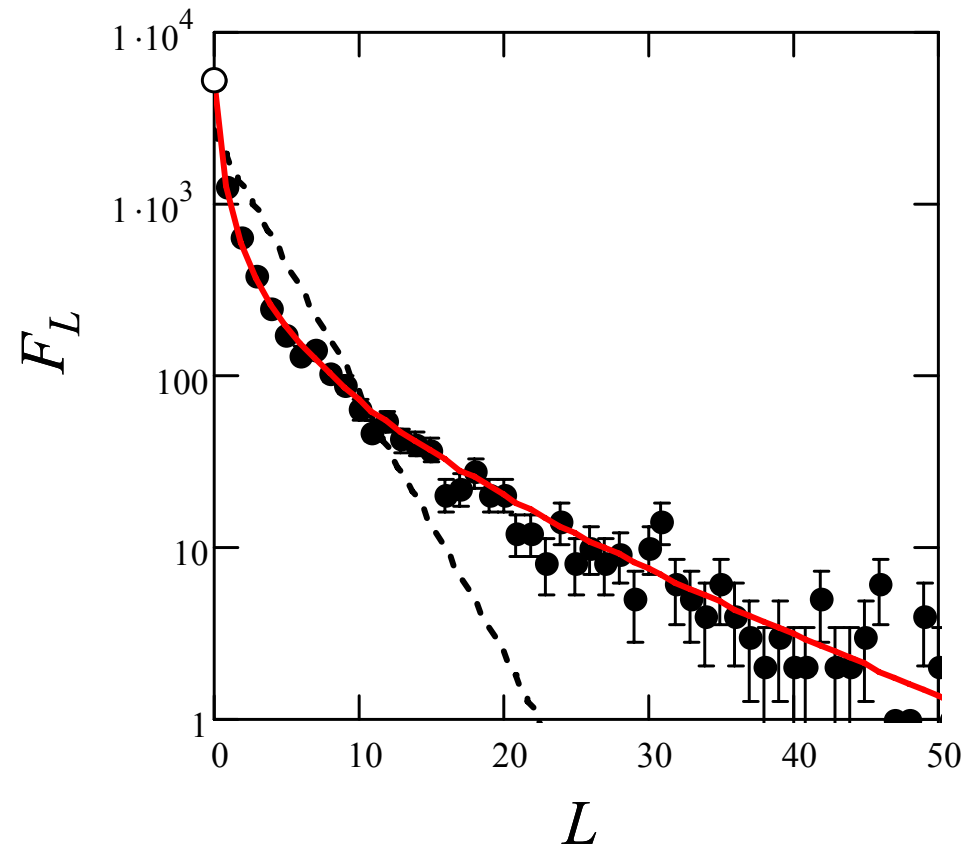

**FIGURE 5.** Burst size distribution $F_L$ at 81 K, obtained by dividing the values in fig.4 by $L$. The value for $F_0$ is shown as an open circle. The curves are explained in the caption of fig.4.

## *Spacing and Length of Bursts*

The time $\Delta\tau$ between bursts shall be the time between the first events of two subsequent bursts. The distribution of $\Delta\tau_m$ for all bursts ($m = 1\ldots M$) is shown in fig.6. The first bin is reduced, since bursts have to be farther apart than their length. The fact that this is an exponential distribution indicates that the bursts occur at random times and are not correlated with each other; the slope of the line is determined by the burst rate $M/t_N$.

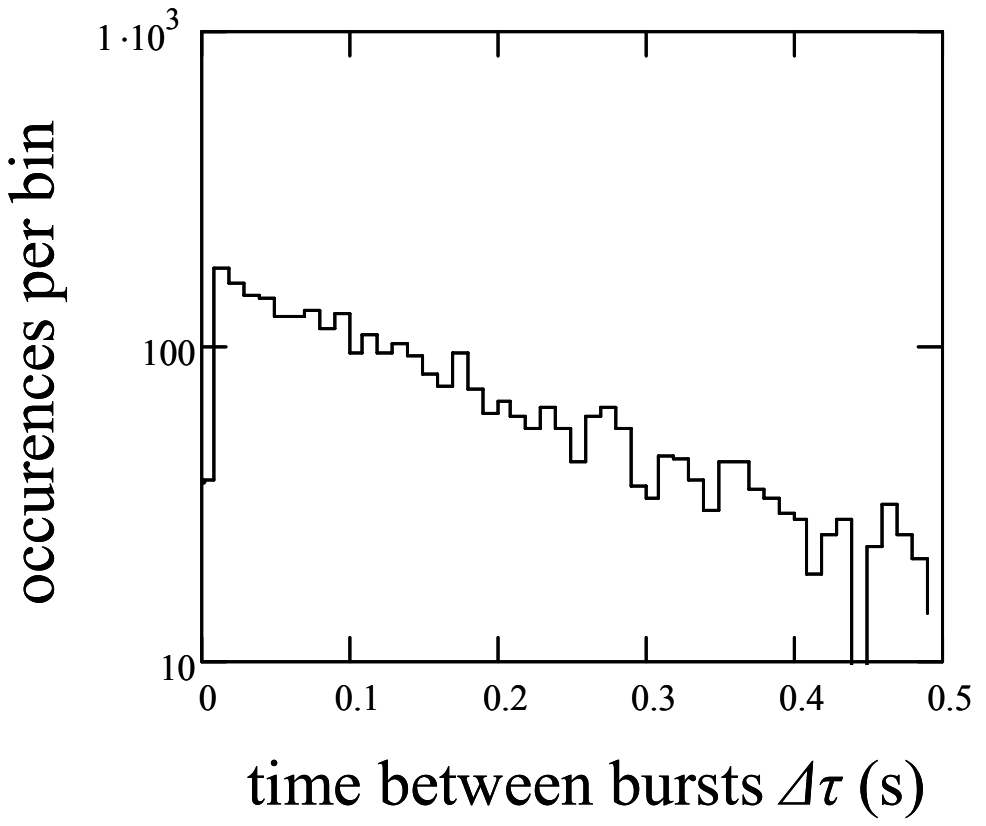

**FIGURE 6.** Distribution of times $\Delta\tau$ between subsequent bursts. The time bins are 10 ms wide.

The burst length (duration) is given by the time between the first and the last event in the burst. The average length of bursts of size $L$ is labeled $D_L$. The distribution of $D_L$ for all bursts is shown in fig.7. It is certainly quite remarkable that bursts may last up to 10 ms, and that there is apparently a mechanism that correlates pulses over such a long time span. It may be of interest that the burst length distribution $D_L$ is, to a good approximation, proportional to $\sqrt{L}$ (solid line in fig.7).

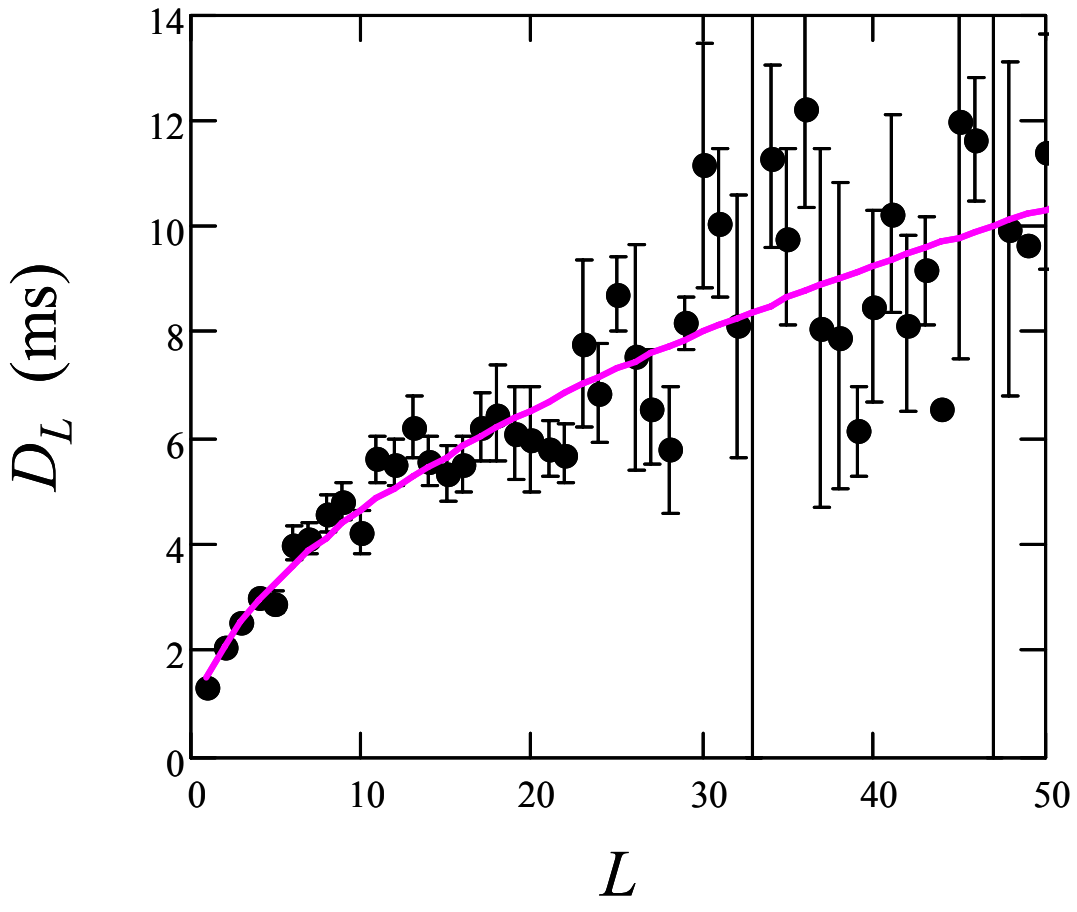

**FIGURE 7.** Average length (duration) of bursts $D_L$ versus $L$.

## *Timing within Bursts*

We now study the time evolution of individual bursts. Let us define $\delta\tau_{m,\ell}$ as the $\ell^{\text{th}}$ interval of the $m^{\text{th}}$ burst, where $m$ ranges from 1 to $M$, and $\ell$ from 1 to $L_m$, the size of the $m^{\text{th}}$ burst. Selecting all bursts of a given size $L$, we plot in fig.8 the corresponding intervals $\delta\tau_{m,\ell}$ versus their order of occurrence, $\ell$, in the burst. Thus, the first column of symbols in the figures contains the first intervals for all bursts of the selected size, and so on. From fig.8 it is obvious that the intervals between pulses in a burst are not random, but that their average is growing, as the burst progresses, from a few to 20 μs in the beginning (shorter for the larger bursts), to about 2 ms (independent of burst size) at the end of the bursts. This slowing down of the emission rate by two orders of magnitude during the evolution of a burst is quite remarkable and must play an important role in discriminating between models for the non-thermal dark rate.

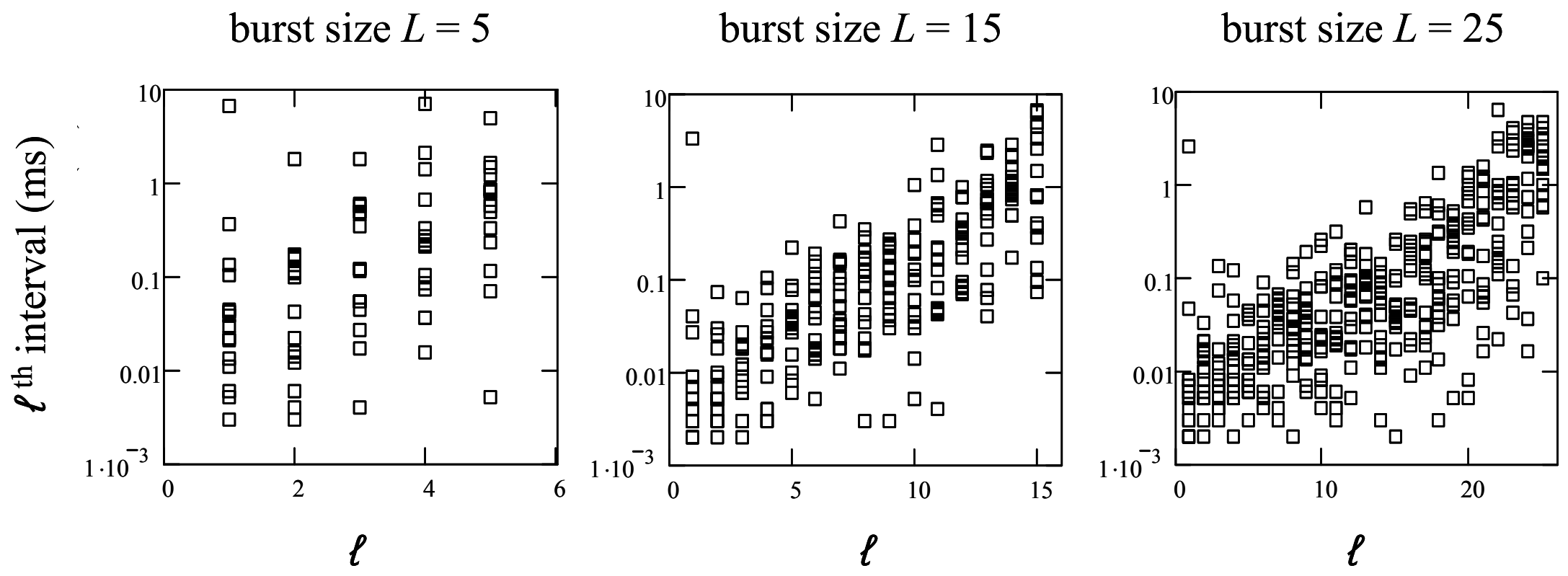

**FIGURE 8.** Time intervals $\delta\tau_{m,\ell}$ for all bursts of a given size $L$ versus the order $\ell$ in which they occur in the burst, for three arbitrarily chosen burst sizes.

Earlier we have mentioned that the non-thermal dark rate is not sensitive to the PM operating voltage. To make sure, we have carried out the analysis described in this section with two data sets acquired with PM voltages of 1650 V and 1850 V. There was no discernible difference between the results of the analysis for the two cases.

## Burst Temperature Dependence

The frequency distribution for the pulse spacings $\Delta t$ for a measurement at 296 K is shown in fig.9, which should be compared to fig.3. It is evident that there are still excess events at short intervals, however there are a lot fewer of them (10%). The corresponding burst size distribution at 296 K is shown if fig. 10 (analogous to fig.5).

We have pointed out earlier that the grouping of pulses is fully described by the number of bursts $M$ (or the burst rate $M/t_N$), and the average burst size $<L>$. These two parameters are shown as a function of temperature in fig.11. Apparently, the increase of the dark rate with decreasing temperature is due to an increase of the burst rate as well as the burst size.

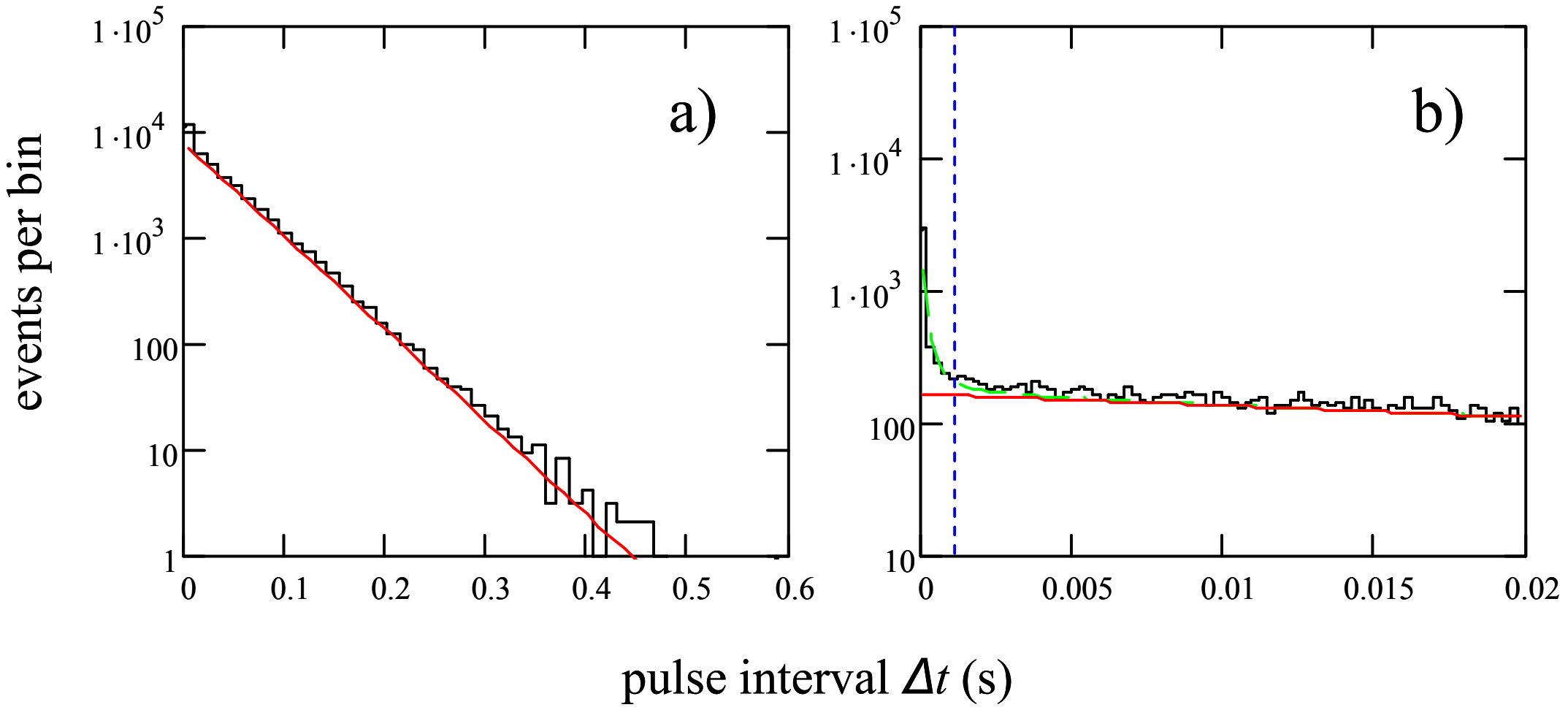

**FIGURE 9.** a) Distribution of $N$ = 40384 intervals $\Delta t$ observed at $T$ = 296 K. The time bins are 12 ms wide. The red solid line is an exponential fit, starting at 0.04 s.
b) shows fig. a) on an expanded scale. The time bins are 250 μs wide. The dashed blue line divides the data into 'short' and 'long' intervals. There are $N_s$ = 3969 short intervals ($\Delta t$ < 0.7ms).

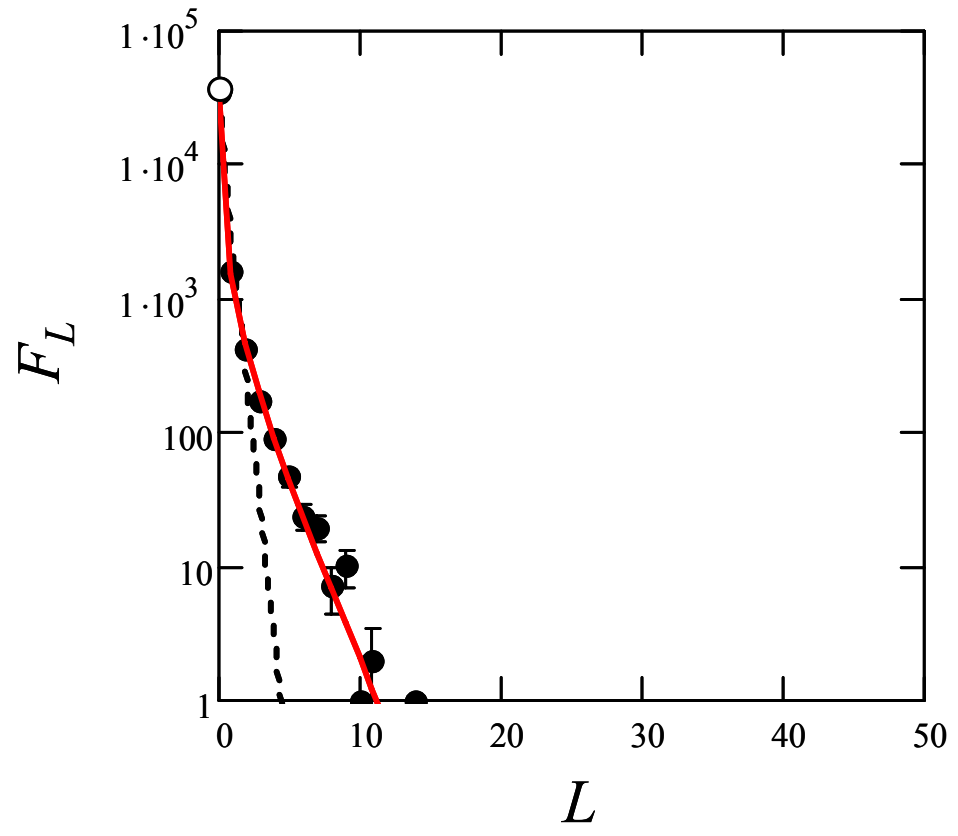

**FIGURE 10.** Burst size distribution $F_L$ at 296 K. The value for $F_0$ is shown as an open circle. The total number of bursts is $M$ = 2350, the average burst size is $\langle L \rangle$ = 1.69. The solid line is calculated from eqs. 3 and 4, using these values. The dashed line shows the distribution that would result if the same number of short and long intervals were arranged at random, using $p$=0.098 for the fraction of short intervals (see caption fig.9).

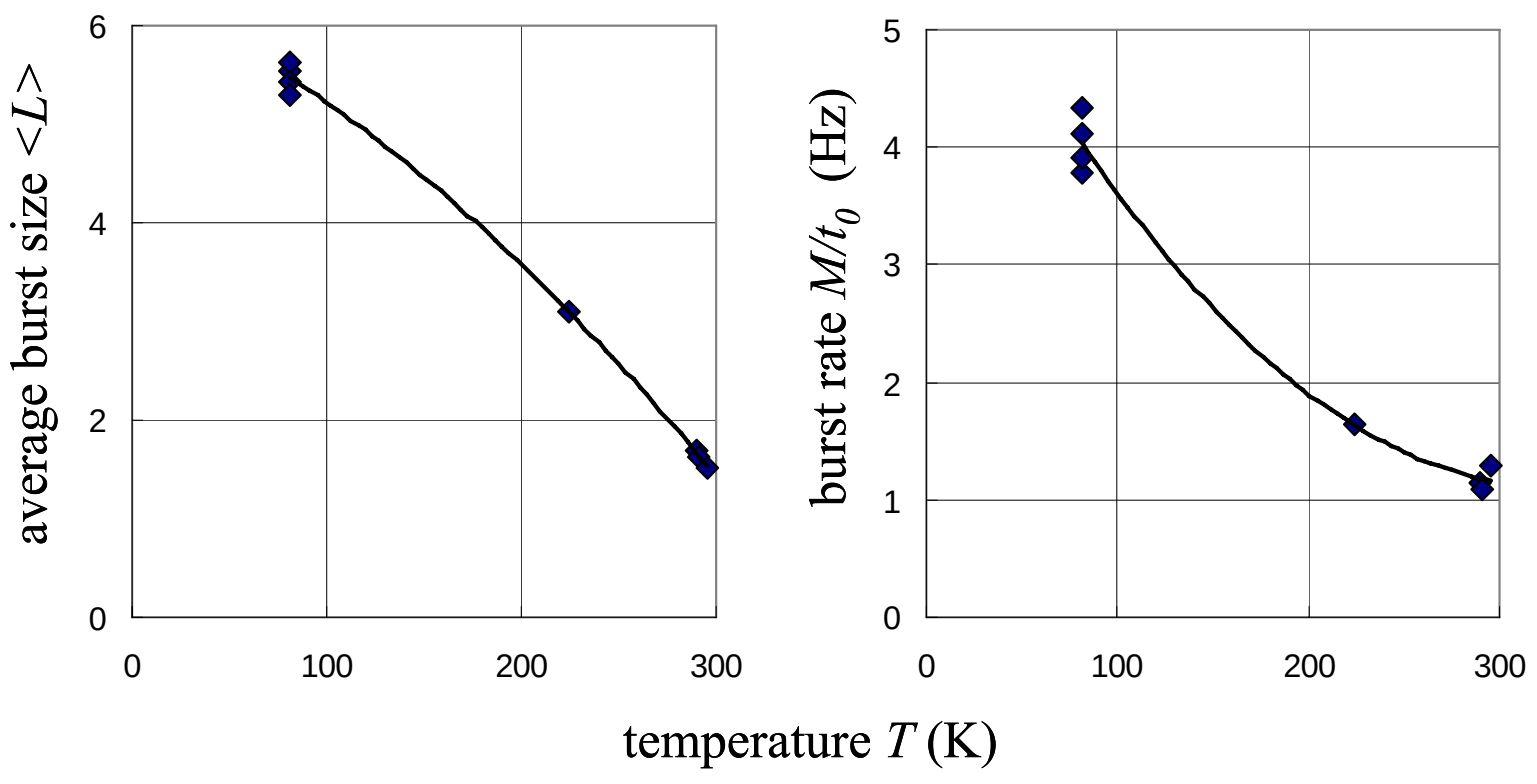

**FIGURE 11.** Average burst size <*L*> and burst rate as a function of temperature *T*. The symbols represent measurements taken at different times. The lines are to guide the eye.

# SUMMARY

The spontaneous pulse response of a photomultiplier in the absence of light has been studied as a function of temperature. At room temperature the dark rate is dominated by thermionic emission. When cooled below room temperature, the pulse rate initially decreases, but then turns around near 250 K and continues to rise all the way down to 4 K. This phenomenon has been called the non-thermal dark (NTD) rate. The main features of the NTD emission are the following:

- For bialkali cathodes on a Pt layer below 200 K (which includes most of the available data), the NTD rate
  - is proportional to the cathode area (→ eq.2),
  - increases exponentially with falling temperature (→ fig.1 and eq.2).
- The onset of the NTD effect is seen also in tubes without a Pt underlayer, and with different cathode materials.
- NTD events are single electrons emitted from the photocathode.
- The NTD rate is not much affected by the electrical field at the cathode surface (→ fig.2).
- The NTD events tend to occur in bursts (with *L* events) with the following properties:
  - the number of events in all bursts of a given size *L* is an exponential function of *L*. (→ eq.3),
  - both, the average burst size and the burst frequency increase with decreasing temperature (→ fig.11),
  - bursts occur independent of each other at random times (→ fig.6),
  - the length of bursts may be as large as 10 ms; the average burst length is proportional to $\sqrt{L}$ (→ fig.7).

- as a burst evolves, the time intervals between subsequent pulses increase from a few μs to about 2 ms. (→ fig.8)

The features of the NTD effect are not explained by known dark current mechanisms, such as field emission, radioactivity or Čerenkov light from cosmic rays. Therefore, we believe that NTD emission is not yet understood. It is hoped that the present set of data may serve as an encouragement to look for an explanation and as a test of theoretical models as they emerge in the hopefully near future.

The practical implications of the data shown here depend on the particular experiment. If there is enough light available to discriminate above the amplitude for a single photoelectron, the NTD counts are largely suppressed. If the dark counts are part of the measurement, however, background subtraction is made more difficult by the burst-like structure of the dark counts.

## ACKNOWLEDGMENTS

I would like to thank D. Baxter, T. Rinckel, T. Sulanke, D. Sprinkle, and my friends in the machine shop at the Indiana University Physics Department for their support of this work.